# Sunspot Numbers from ISOON: A Ten-Year Data Analysis

K.S. Balasubramaniam,
Space Vehicles Directorate,
Air Force Research Laboratory, Kirtland AFB, NM, USA
Email:.bala@yahoo.com

T.W. Henry
Boston College & Space Vehicles Directorate,
Air Force Research Laboratory, Kirtland AFB, NM, USA

**Abstract**: Sunspot numbers are important tracers of historical solar activity. They are important in the prediction of oncoming solar maximum, in the design of lifetimes of space assets, and in assessing the extent of solar-radiation impact on the space environment. Sunspot numbers were obtained visually from sunspot drawings. The availability of digital images from the US Air Force Improved Solar Optical Observing Network (ISOON) prototype telescope concurrent to observer-dependent sunspot numbers recorded at the National Solar Observatory (NSO) has provided a basis for comparing sunspot numbers determined from the two methods. We compare sunspot numbers from visual and digital methods observed nearly simultaneously. The advantages of digital imagery are illustrated.

Keywords: Sunspots, solar cycle, sunspot numbers

.

1. Introduction

The temporal fluctuation of the daily sunspot numbers is the underlying basis for understanding and modeling the historical nature of solar activity, which in turn drives the prediction of solar cycles (*e.g.* Hathaway *et al.*, 1999; Hathaway, 2010). Counting sunspots is fraught with inaccuracies (see, *e.g.*, Hoyt, Schatten, and Nesme-Ribes, 1994; Cliver, Clette, and Svalgaard, 2013). A complex system of **consistency factors** continues to drive the determination of the scaled international sunspot index [$R$], including telescope optical quality, resolution, atmospheric seeing, photometric sky quality, and observer vision. These consistency factors have changed over time due to the fact that sunspot observing is a human activity. Hence, attempts to normalize our measurements of sunspot activity over decades and centuries that are obtained from different data sources lead to unintended errors. In addition, determining the boundary of a sunspot region and to which region the observed sunspot or pore belongs has led to a questioning of the authenticity of the very definition of a sunspot index. However, despite the absence of consistent and verifiable data, it is vital to capture a model variation of the solar cycle over centuries. The international sunspot number is the most viable measure of inference to represent solar cycles. We refer the reader to the body of work as captured in the series of Sunspot Number Workshops organized by Cliver, Clete, and Svalgaard (ssnworkshop.wikia.com/wiki/Home) to help understand the difficulties of reconciling sunspot numbers.

The solar physics literature has numerous references to a rich tradition of research efforts to compare sunspot numbers from various observatories. To illustrate, we refer the reader to additional efforts to include sunspot number recordings from digital and or historic data. They include measures from Debrecen Solar Observatory (Györi *et al.,* 2010), Kanzelhöhe Observatory (Pötzi *et al.,* 2016), Kislovodsk (Tlatov, 2014) and Kodaikanal (Sivaraman *et al.,* 1992).

### *Sac Peak White-light Telescope*:

Historically, observers at the National Solar Observatory (NSO), and its predecessor the Sacramento Peak Observatory have kept daily records of sunspot number [$R$] made from hand-drawings from 1953 – 2006. A 15-cm telescope projected a white-light image (no filters) on a white-board using an eyepiece to form a 19-cm image. The eyepiece was manually adjusted every day to fill the visually focused solar image so that the visual limb was encircled within the 19-cm circle. (see: http://nsosp.nso.edu/node/16). The sunspot number [$R$] (also known as the Wolf number or the International Sunspot Number: ISN), is defined by

$$R = k\,(10\,g + f) \qquad (1)$$

where *k* is an observatory-dependent factor, *g* is the number of groups, and *f* is the number of individual spots observed.  For Sacramento Peak images, *k* was always assigned a value of unity, for unknown historical reasons.

These indices and the corresponding daily drawings were transmitted to the NOAA Space Weather Prediction Center and its predecessor at Boulder, CO, USA to contribute to the international sunspot number.

*ISOON Telescope*:

A partially overlapping and nearly decade-long (2002 – 2011) digital effort to capture a consistent verifiable measure of sunspot index was made using the Improved Solar Observing Optical Network (ISOON) prototype telescope operated at the NSO at Sunspot, New Mexico, USA, by the Air Force Research Laboratories.

The ISOON telescope is a 25-cm aperture, polar axis, evacuated refractor with a 5000 mm primary focal length.  The spectral filters consisted of 150 mm aperture dual Fabry-Perot filters with a band-pass of 0.1 Å, at 6303.15 Å, a region of the solar spectral pure continuum, devoid of any spectral line.  The detector was a 2048 × 2048 pixel water cooled XEDAR CCD camera with 4096 grey levels.  Continuum images at an angular resolution of 1.1″ per pixel were acquired at a cadence of five minutes. The raw images are scaled to a solar radius of 890.4 pixels and de-rotated by the solar position angle, $P_0$. (Neidig *et al.*, 1998; Balasubramaniam and Pevtsov, 2011)

One of us (TH) was the same observer who compared the sunspot groups from both hand-made drawings and the digital images for a period of a year and a half during 2003 – 2004, from these two different instruments.  This article describes the insights gained from the comparison of ISOON data with historic Sacramento Peak sunspot data.

. In Section 2 we describe the data from the individual instruments and inter-compare the individual measures. In Section 3 we describe the insights gained from analyzing the long-term and short term variations.

2. The Data

In 2003 – 2004 a 17-month study was conducted comparing hand drawings from the NSO's white-light telescope with ISOON continuum images. The observing sites were approximately 100 feet apart.

2.1 Historic Sunspot Data

At the 15-cm telescope equipped with a projection board, observers sketch a sunspot drawing on a sheet of paper on which the white-light Sun is focused. Sunspots are drawn with a pencil. The paper is usually of high quality so as not to confuse dark flaws in the paper with solar features or sunspots. Typically, penumbrae are drawn with a sharp pencil while umbrae are drawn and filled in. A second piece of paper may be waved in the light path in front of the drawing to ensure a particular dark feature being projected is actually of solar origin. White-light faculae may also be added to the drawing. In the next step, the sheet is removed from the projection board, groups are marked, the Wolf number is computed, labels are added, and the drawing is complete. An example of the sunspot number drawing is shown in Figure 1. The results from the (hand-drawn) sunspot number were sent to NOAA since 1953 daily, although the data-accumulation process started in 1948. The data was sent by facsimile. (Note: This method of sunspot counting is the basis for sunspot-number measures used by USAF's Solar Electro-optical Network, whose data from the Solar Observing Optical Network (SOON) is transmitted to NOAA and then to the World Data Center for Solar Terrestrial Physics, which helps in reconciling international sunspot numbers)

The method used to count sunspots at the NSO was handed down from observer to observer. Each umbra was counted as a separate spot; a penumbra without an umbra is neglected when present within a spot group that has a different umbra to contrast with. If a separate penumbra from a region on the limb can be identified, which is not a part of any other spot group, then that is counted as a spot group. This information was traditionally handed over from observer to observer. Typically, the drawing was compared to a NOAA Solar Region Summary to separate and determine the number of groups when labels are applied.

2.2 Digital Data from ISOON

When the USAF's ISOON prototype was constructed, the ability to make automated measurements of true continuum images was incorporated into the system (Neidig *et al.*, 1998; Balasubramaniam and Pevtsov, 2011). The method is as follows: i) the solar disk image is corrected for atmospheric and optical distortions and is made circular. ii) A radial averaging function is used to remove penumbrae, spots, and faculae. We used the criteria that the local intensity (averaged to unity) is between 0.92 – 1.08 to define the quiet Sun radial profile using a polynomial fit. iii) A 2D limb-darkening function is reconstructed. (Tests of the limb-darkening accuracy have been verified over a few thousand images). iv) The limb-darkening function is subtracted from the solar disk image. Removal of limb darkening helps to identify individual smaller spots that might otherwise be missed, and to identify the location of faculae with better visual contrast. The resulting image is also used to calculate irradiance deficit. (The quiet-Sun solar disk has an approximate value of unity, with embedded granular contrast variations accounting for the fluctuations). v) Contiguous penumbral areas are identified where the local intensity is below 0.92, and umbral areas, where the local intensity is below 0.68, are identified.

We elaborate on how we identified the sunspots on an image that has been corrected for limb-darkening effects. Three simultaneous criteria are used to determine a feature (umbra or penumbra). The criteria are based on intensity, area, and temporal continuity, as follows:
(a) Intensity: The normalized intensity values of 0.92 and 0.68 as cut-offs have been determined by using intensity distributions in a histogram. Examining the inflexion points in a histogram we have determined the intensity cut-offs. The details of this technique are similar to that applied by Balasubramaniam (2002). The consistency of the cut-offs were determined by examining its application to identifying umbral and penumbral boundaries with ISOON data, obtained over a year in 2003.

(b) Area: A feature (umbra or penumbra) area is determined by the following method, adapted from Smith and Smith (1963) and reported by Henry (2015). A pixel on the image (1.1″ square) defines a grid unit. The solar radius [*r*] in each image is 890.43 pixels. A contouring algorithm determines the boundary of a feature, within the solar image. The vertices of the boundary are used to measure the area it encloses, in grid units. The true area, in micro-hemispheres, is given by:

$$A = Ac \frac{10^{-6}}{As} \qquad (2)$$

Here Ac is the area of the feature and As is the area of the sun, in grid units. The exponent refers to the micro-hemisphere. By trial and error we have determined that the number of vertices has to be at least 12 in order to eliminate noise and dust spots in the image. The smallest unit area measured is one solar micro hemisphere.

(c) Temporal continuity: For a feature (umbra and/or penumbra) to be recognized automatically, it has to be present in three consecutive images. Since the cadence of each image is at every five minutes, the image has to be present for at least 15 minutes. This eliminates granular lanes, dust and short-lived pores. The continuity is determined by projecting the contours of each image onto the contours of the next image. Should the two contours overlap, the feature is accounted.

Images are examined visually for transient dust specks or shadows. The groupings of active regions are manually delineated against the current day's NOAA region summary report, issued at 00 UT. The resulting images were then fed into an algorithm to measure sunspot numbers, umbral and penumbral areas, and irradiance measures (Neidig and Henry, 2004). Human intervention was necessary to allow for a visual recognition of what constituted the assignment of pores to an active region, potential specks of transient dust, and errors due to automated flat-fielding of CCD images when clouds interfered with the acquisition of flat-fields.   The right-hand side of Figure 1 illustrates a comparative digital image.

 It is important to note that the digital images have been corrected for dark-current residuals and flat-fielding effects. In addition, to normalize the intensity of sunspot features irrespective of where they are on the disk, the images have been limb-darkening corrected (*cf*. Figure 2).   Even on the digital images, spot-groups were visually identified against NOAA group numbers, and where NOAA numbers did not exist they were assigned a temporary region number.   The automated program recognizes the contours, counts, and locations of all penumbrae and umbrae based on thresholds discussed above.   The resulting image is also used to calculate the irradiance deficit, the relative darkness of spots and the areas of sunspot penumbra and the umbra.

The system is designed to take the NOAA Region Summaries and provide a sunspot number count automatically, as is shown in Figure 3. The labeling is sometimes altered manually to prevent overlap or visual distractions.  The data-processing algorithm can obtain sunspot number counts and irradiance deficit irrespective of the presence of an operator. However, human intervention is necessary when sunspot groups split or new groups appear in the vicinity of older groups, or labels appear to overlap that causes confusion upon a visual examination.   Even seeing conditions are measurable from the image quality.   Once initiated, the algorithm will run continuously, providing Wolf number, area, irradiance deficit, and an equivalent drawing without further intervention.

2.3 Comparing Hand-Drawings *versus* Imagery

An important part of transitioning from a 52-year hand-drawn data base to a digital form was to ensure the consistency of the digital data overlap with the hand-drawn data. Table 1 shows the results of measurements from sunspot drawings and ISOON imagery for the start of the comparison in 2003. This comparison was continued until 11 May 2004. NSO hand-drawings were permanently discontinued after this date, driven by programmatic decisions. The Table 1 in itself will not show the smaller pores; however, the digital imager detects smaller and fainter penumbrae and pores, reflected in the G-number, whereas the drawing identifies more individual umbrae, as reflected in spot-counts. An immediate conclusion from this analysis is that a visual observer easily misses fine spots and spot groups, particularly near the limb. Equation (1) has a group factor of 10, and hence finding additional smaller sunspot groups increases the sunspot number [$R$] by at least by 11 counts per group One should note that seeing impacts counting. If poorer seeing blurs the existence of spots, then the numbers are lower, similar to a visual observer losing sunspot counts. This poses a difficulty in reconciling older sunspot numbers from across multiple telescopes that are observer-dependent, while bringing them to the modern digital age. The advantage of the digital technique is that since data are preserved one can verify the authenticity of the sunspot counts, unlike the hand-drawn sunspot drawing records.

2.4 Comparing ISOON Measures to International Sunspot Numbers

It is important to compare the measures of ISOON sunspot numbers to the ISN (http://www.sidc.be/silso/datafiles) which serve as historical reference data. We compare the two sunspot numbers as shown in Figure 4. The updated ISN numbers (corrected since July 2015) are higher than the corresponding numbers as recorded by ISOON, by a factor of ≈1.3. The correlation coefficient of ≈ 0.95 between the two data sets, for the time period investigated.

A high (≈ 95 %) correlation between the two numbers shows what high confidence we must attribute to the international sunspot numbers when digital data were historically absent. The curve provides an opportunity to revert digital numbers to international numbers using the linear equation:

$$R_{Int} = k\ R_{ISOON} + B \quad (3)$$

where *k* = 1.3 for the comparison period (May 2003 – March 2012). To understand the extent of the differences between ISOON and ISN during nearly a decade (2002 to 2011), we show the actual differences in Figure 5. Notice that during high solar activity, ISN systematically overestimates the number of sunspots. These are particularly noticeable around solar maxima years at 2004 and 2012. This immediately leads to a consideration of how the *K*-factor (an annual reconciliation index) varies with time. Table 2 shows an annual variation of the *K*-factor (see Equation (3)). The solar-maximum periods show a high correlation, while the solar minimum shows a relatively weaker (still high) correlation. The reason is that during times of minimal solar activity, one observatory might see a short lived pore (counted), whereas another observatory will not report this. For example on 13 October 2003, two pores in the south west quadrant were observed late in the afternoon, counted as one spot group. On 14 October there were three pores (in two different groups) seen for a short period of time. The ISOON data did not detect any spots (due to intensity thresholds; the spots were too small or too faint) in the morning on both days, where the images show no pores on the disk. Notice that the ten-year mean correlation coefficient is high because there are a disproportionate number of points (422 days) during which there were no sunspots during the solar minimum. These results contrast the subtleties of sunspot counts from figure drawings, when compared to digital data.

Despite the high correlation, other sources of difference include that while the ISOON number is determined from a "snapshot" of the sun over a 15-minute period of time, the ISN numbers have been reconciled from about 20-observatories drawn from a 24-hour period. This would result in differences due to either counting short-lived spots during ISOON observations or completely missing them, which was accounted for by the ISN data.

Also not traceable are potential uncertainties contributing to the International Sunspot Number counts due to the presence of transient spots (with lifetimes of a few hours) during a time when an observer records visual sunspots by drawing them. The advantage of digital data is the potential to study the temporal evolution of sunspot numbers for a particular active region, which can provide additional insights into the evolution of solar activity within short periods of time. This correlation demonstrates that there needs to be an additional reconciliation factor of the new ISN that is a function of solar cycle phase.

2.5 Comparing Pre-2015 ISN to Post-2015 ISN to ISOON Measurements

The Sunspot Index and Long-term Solar Observations website at the Royal Observatory of Belgium (www.sidc.be/silso/datafiles) changed the original ISN series (version 1.0; also referred to as "OLD", hereon) to a new ISN series (version 2.0; referred to as "NEW", herein). Reported in this article, thus far, is the comparison of digitally derived ISOON Sunspot Numbers with the new series, 2.0. From the stand-point of ISOON data, it will be useful to compare the difference between the OLD, and the NEW ISN values. To help us understand this difference in Figure 6. The figure clearly shows that there is a continued non-linear trend that increases with the solar cycle. One should bear in mind that the ISN numbers (OLD or NEW) are reconciled from a large number of observatories around the world and irrespective of how the trends in the OLD versus the NEW data sets are reconciled; the visual counting of sunspots become increasingly different in amplitude, with the solar cycle, i.e. underestimated using the OLD ISN and overestimated using the NEW ISN (see Figure 5). The correlation coefficient has changed insignificantly irrespective of the source of the ISN. They are: 0.956 for ISOON number versus OLD, 0.957 for ISOON number versus NEW.

The insignificance of the difference between OLD or NEW ISN numbers when compared to ISOON numbers during the time-period of 2002 – 2012 is further illustrated in Figure 7, where we depict a scatter plot of the difference, referenced against ISOON number count. Here we once again see that the difference in the sunspot counts is exacerbated during higher solar activity or larger sunspot numbers.

2.6 Irradiance Reduction due to Sunspot Blocking

An additional advantage of digital imagery is that it provides a measure of reduction in continuum intensity due to the blocking of sunspots (Neidig and Henry, 2004). The reduction, or deficit [$f_\lambda$], in irradiance due to a dark feature with area $A$ [pixels$^2$] and intensity [$I_\lambda$] is

$$f_\lambda = 6.18 \times 10^{15} \text{ cm}^2 \text{ pix}^{-2} \ (I_\lambda(0)/600) \ I_\lambda \ dA \ /(1 \text{ AU})^2 \ \text{ erg s}^{-1} \text{ cm}^{-2} \text{ Å}^{-1} \qquad (4)$$

where $I_\lambda$ (0) is the intensity of the quiet Sun at disk center, averaged over an area of 110 arc seconds squared (100 × 100 pixels). Note that $I_\lambda$ will appear as a negative number in the ISOON reductions. For λ = 6303.15 Å, $I_\lambda$ (0) = $I_\lambda$/ 0.83 = 3.036 × 10$^6$ erg s$^{-1}$ cm$^{-2}$ Å$^{-1}$ sr$^{-1}$ (Allen, 1991). The intensity [$I$] here is taken from a limb-darkening subtracted image. The number 600 is because ISOON's data camera acquisition system sets the disk-center intensity to 600 counts, as a reference. Figure 8 shows the irradiance reduction derived from ISOON data (2003 – 2011; 1532 days of observing) compared to the area of sunspots as determined by thresholds previously mentioned. The linearity between the two quantities is notable. Perhaps sunspot areas or irradiance deficit measures can be a quantitative measure of the solar activity cycle.

3. Temporal Variations of Sunspot number, Area, and Irradiance Deficit.
   3.1 Long-Term Variations

Having established the basis of the digital measures of sunspot number, area (in millionths of the solar hemisphere) and irradiance deficit, we next explore the temporal variations for 2003 – 2010. Figure 9 shows the solar-cycle dependence of the sunspot number, sunspot area (penumbral area + umbral area), and the irradiance deficit. Within each plot is an inset that shows the prior 30-day data on 09 April 2010. The label

"current day" on the plot simply reflects the fact that this display can be updated at any given time when new observations accumulate. The solar-cycle behavior is evident in the comparison, which starts in 2003 at the end of the maximum in Cycle 23 and ends in 2010 at the beginning of the new Cycle 24. We see that the sunspot number is not tightly correlated with sunspot area. The clarity of the sunspot area increase, and the corresponding increase of the irradiance deficit during the end of October 2003, when the Sun was exceptionally covered with a large number of sunspots, is evident. Similarly the inset figures show the 28-day sunspot rotation off the solar disk and the dramatic changes in sunspot area and irradiance deficit during late March to early April 2010.

3.2 Intra-Day, Short Term Fluctuations of Sunspot Component Areas During Flares

In attempting to understand the short-term variations of these quantities' influence and to help attribute the underlying physics to these changes, it is instructive to see how sunspot component areas (namely umbra and penumbra) change within a day, particularly when a solar flare occurs. With the availability of consistent high-resolution digital data at a far higher cadence (one minute or higher) the contrast of intra-day changes compared to daily changes becomes clear. Figure 10 illustrates the intra-day changes. The panels in Figure 10 are for the X6.5 flare of AR 10930 on 06 December 2006. The mean H$\alpha$ intensity for the flare, averaged over the active region, is shown on the top right panel to help in identifying the start of the flare (vertical line). This figure illustrates that intra-day changes in sunspot activity are well captured by area and intensity changes, rather than sunspot number measures, afforded by high quality digital images.

The complexity of sunspot dynamics can be represented by changes in the umbral and penumbral intensities and areas. For reference, the relative H$\alpha$ light curves of the activity are plotted. The H$\alpha$ light curves intensities are relative to a quiet-Sun solar chromosphere at the disk center. The H$\alpha$ light curve is a measure of the chromospheric activity, similar to the GOES X-ray light curves, except that H$\alpha$ measures the changes in activity in individual active regions. The mean umbral and penumbral continuum intensities are derived after the umbra and penumbra are delineated by a contoured threshold as described, earlier. The umbral and penumbral areas are distinct and in units of millionths of the solar hemisphere. The vertical lines in the figure show the start of the solar flare times for the active region as documented the NOAA Solar Activity Summaries. For example, we see that after the 06 December 2006 X6.5 flare the umbral areas decrease and the penumbral areas increase. A possible interpretation is that the horizontal penumbra fields have become relaxed and resulted in vertical umbral fields. Similarly the amplitude of the umbral intensity and area being lower before the flare and after the flare can be construed as the suppression of umbral oscillations by stressed magnetic fields. However such conjectures need to be fortified by statistical measures from repeated flaring states.

Hence, digital imagery offers the added advantage of measuring intra-day area changes tracking the rapid changes in solar activity that can be retroactively verified. Such measures strengthen our understanding of changes in sunspot numbers over shorter time-scales than the original once-a-day sunspot number-counts.

**Discussion and Summary**.

We have demonstrated the advantages of using consistent digital sunspot images and the insights gained from analyzed sunspot numbers from both sunspot drawings and digital data. We have established a coherent comparison of sunspot numbers from both sources, and trace potential errors when comparing international sunspot numbers to recent digital data. We have shown that when comparing the pre-2015 ISN data (OLD) to the post-2015 ISN (NEW) data, there is a consistent trend of ISN numbers increasing with increased solar activity, when compared with digital imagery data. The only contrast between the OLD and the NEW data is that is that they differ by a factor of $\approx 0.6$ We have also shown the advantage of intra-day variation of sunspot areas as an alternative to sunspot numbers where digital data are available. We have established the conversion metrics to go between sunspot numbers and sunspot-area measures, as needed in the context of the research.

The merging of digitized continuum images with projection-board drawings as a means to compute Wolf numbers provides a better picture of solar activity. One can characterize what was seen on the projection board. To understand and develop an effective conversion between the two methods under similar conditions, digital images need to be acquired at nearly the same time as projection board drawings, at similar, if not identical, locations. This article demonstrates such an effort. A projection-board drawing requires an observer to be sitting at the telescope drawing and counting sunspots. If one is unavailable, the opportunity is lost. Any drawing made is not reviewable, although it may be compared with other observations by other observers elsewhere and at different times. In contrast, a digital image can be reassessed as its analysis can be adjudicated at any time in the future.

Sunspots can emerge or dissipate at any point in an observing period. Often faint spots can be seen in an image, persist from image to image and yet be below threshold. Fading in and out, they may or may not be seen on a projected image. The question arises whether to ignore these spots or count them in some way. A researcher needs to be cognizant that counting sunspots may drastically increase the $R$-number, and they must use the results appropriate to the context. One such example was on 13 October 2003, when the south west part of the Sun, close to the disk center, showed persistent visual spots. However, the corresponding digital counts showed no such spots, because they were below the intensity threshold.

In general, imagery from ISOON detects many more spots and groups than ISN suggest, but there are many exceptions. Some of the difference will be in the method of counting spots, identification of penumbra or umbra. However, counting umbrae will greatly exaggerate this difference, which can be alleviated by determining minimum threshold such as in area or darkness to define a sunspot.

Comparison was also made between ISOON spot counts and NOAA/Space Weather Prediction Center counts that are immediately available before reconciliation. SWPC uses the USAF's operational SOON telescope network's projection-board drawings in processing the data. Despite better seeing conditions for ISOON (at high altitude where Sacramento Peak is located), preliminary spot counts reported by NOAA–USAF were consistently greater and often double those seen with ISOON imagery. For example, from 01 January 2008 to 31 March 2011 there were 609 common observations for ISOON and NOAA reports. More than ten percent of the time reports registered more than double the ISOON spot counts.

An image is a means to verify the existence of an active region from an observatory. The method used to assign NOAA active-region numbers can have the effect that active regions may be several days old before being identified. In addition, one must be aware that the NOAA active-region number assignment can be temporally stale by as much as a whole day. In our experience at NSO, numerous one- and two-day events, where new active regions are not numbered, were noted. A one-day event may be understood as observations at NSO in New Mexico are relatively late in the observing day as compared to other observatories around the world. As confirmation of a region requires observations twelve hours apart, a two-day event may be understood as observatory downtime and a region observed by the same observatory over two days before confirmation.

In tracking the intra-day variation of solar activity, we have demonstrated that sunspot-component measures, such as umbral and penumbral areas and intensities, dramatically change during and after large solar flares. The underlying changes in sunspot areas cannot be reflected in sunspot numbers, which are coarser measures of solar activity on time-scales of a day or more. Integrated sunspot areas and irradiance deficits can be diagnostics of alternate measures of solar activity to monitor finer changes. The important point here is that sunspot imagery and quantitative changes are better reflected in measures of component solar activity than those represented by sunspot-number changes, when considering finer representations of solar activity in the context.

If one were to construct a network of telescopes, as was planned for ISOON, issues to be resolved would include standardization of wavelength, filter width, and aperture of a candidate telescope. A high-resolution telescope will see very fine spots and can be expected to see more than the human eye. Integrated over time, a more accurate picture of solar activity can be obtained from a telescope with multiple images than a single observation once a day.

Acknowledgments

This work was supported by the Space Vehicles Directorate, Air Force Research Laboratories (AFRL) and The Air Force Office of Scientific Research, AFRL. We are grateful to Donald Neidig and numerous members of joint AFRL–NSO ISOON Team, the AFRL Space Vehicles Directorate, and the NSO for developing and operating the ISOON prototype telescope for a decade and preserving its data. Particularly, Neidig's insights into calibration of the data and into sunspot counting techniques have inspired this article. We thank Ed Cliver, Frédéric Clette, Leif Svalgaard, and numerous colleagues for sustaining the importance of the sunspot number research through these workshops.

## Disclosure of Potential Conflicts of Interest

The authors declare that they have no conflicts of interest.

**REFERENCES**

Allen, C.W.: *Astrophysical Quantities*, Athlone Press, London 1991.

Balasubramaniam, K.S: 2003, *Astrophys. J.* **575**, 553.

Balasubramaniam, K.S., Pevtsov, A.A.: 2011, *Proc. SPIE* **8148**, id. 814809.

Cliver, E.W., Clette, F., Svalgaard, L.F.: 2013, *Cent. Eur. Astrophys. Bull.* **37**, 401.

Györi, L., Baranyi, T., Ludmány, A.: 2010 in *The Physics of Sun and Star Spots Proc. IAU Symp* **273**, Cambridge Univ. Press, Cambridge, 403. ADS: 2011IAUS..273..403G, DOI: 10.1017/S174392131101564X

Hathaway, D.H.: 2010, *Living Rev. Solar Phys.* **12**, 4. doi: 10.1007/lrsp-2015-4.

Hathaway, D.H., Wilson, R.M., Reichman, E.J.: 1999, *J. Geophys. Res.* **104**, 22375.

Henry, T.W.: 2015, *Solar Ephemeris Applications and Image Analysis*, MS Thesis, Texas Tech University, Lubbock, TX,

Hoyt, D.V., Schatten, K.H., Nesme-Ribes, E.:1994, *Geophys. Res. Lett.,* **21** (18), 2067.

Hoyt, D.V., Schatten, K.H.: 1998, *Solar Phys.* **179**, 189. ADS: 1998SoPh..179..189H, doi:10.1023/A:1005007527816.

Neidig, F., Henry, T.: 2004, Comparing Sunspot Numbers Between NSO and ISOON: NSO/Sac Peak Projection Board Drawings vs. ISOON Semi-Automatic Procedure" NSO Internal Technical Document. (Private communication).

Neidig, D.; Wiborg, P., Confer, M., Haas, B., Dunn, R., Balasubramaniam, K.S., Gullixson, C., Craig, D., Kaufman, M., Hull, W., McGraw, R., Henry, T., Rentschler, R., Keller, C., Jones, H., Coulter, R., Gregory, S., Schimming, R., Smaga, B.: 1998, In: Balasubramaniam, K.S., Harvey, J., Rabin, D. (eds.) *Synoptic Solar Physics*, **CS-140**, Astron. Soc. Pac., San Francisco, 519.

Pötzi, W., Veronig, A.M., Temmer, M., Baumgartner, D., Freislich, H, Strutzmann, H.: 2016. *Solar Phys.* DOI: 10.1007/s11207-016-0857-6

Smith, H.J; Smith, E.P.: 1963. *Solar Flares*, Mcmillan, New York.

Sivaraman, K.R., Gupta, S.S., Howard, R.F.: 1993, *Solar Phys.* **146**, 27. ADS: 1993SoPh..146...27S, doi:10.1007/BF00662168.

Tlatov, A.G., Vasil'eva, V.V., Markova, V.V. Otkidychev, P.A.: 2014, *Solar Phys.* **289**, 1403. ADS: 2014SoPh..289.1403T, doi:s11207-013-0404-7.

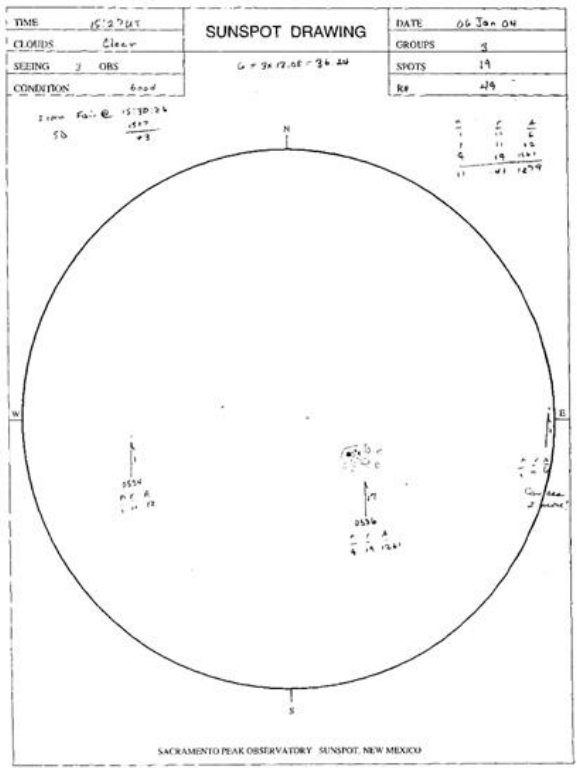 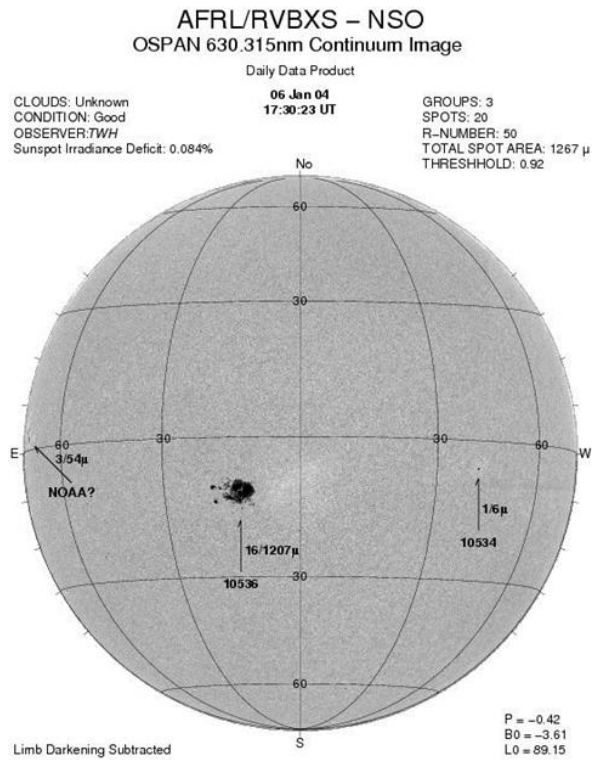

**Figure 1:**. *Left*: Sunspot drawing used to record sunspot numbers for 06 January 2004 using the NSO telescope at the Hilltop Dome. There are 3 sunspot groups and 19 spots, resulting in an *R* number of 49. *Right*: The corresponding digital image from ISOON. The *R*-number is 50. Note that the E and W directions are flipped in the two drawings. The original sunspot drawing is as observed.

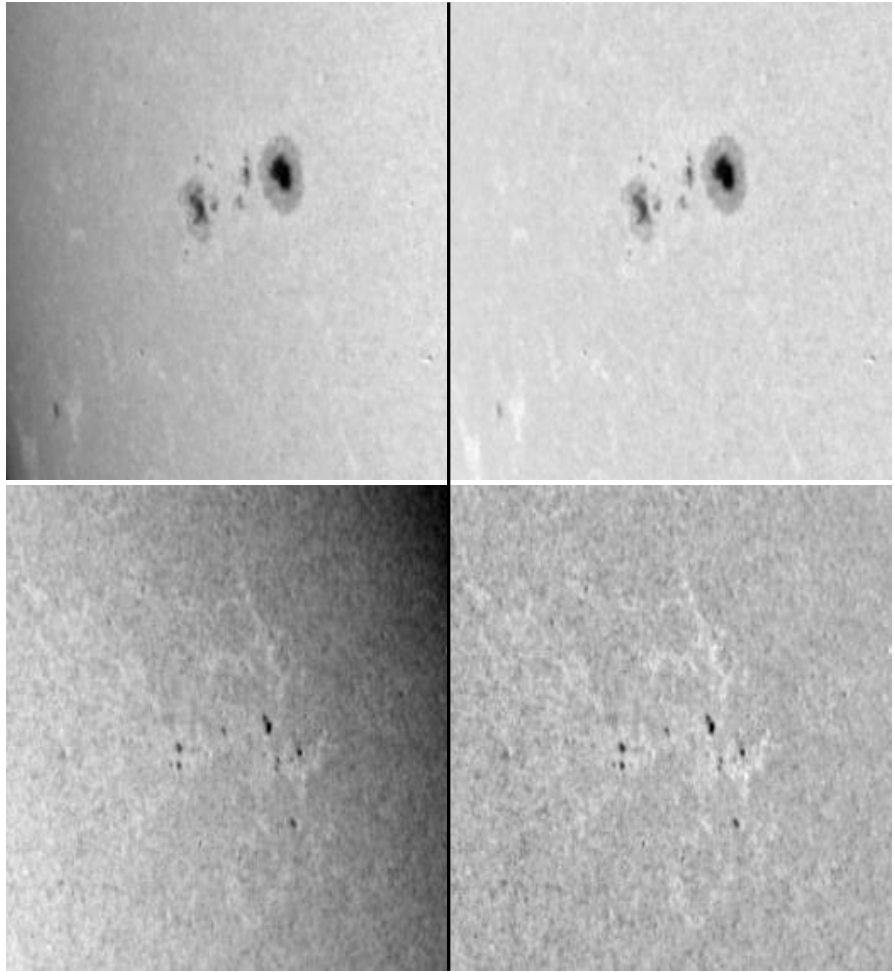

**Figure 2:** . *Top Left*: East-limb example: A continuum image on the east limb with limb-darkening. **Top Right** The same region as the top-left image with limb-darkening removed. **Bottom** West-limb example: similar to top figures with and without the subtraction of limb darkening. The images are ≈ 240 × 290 arcseconds

**Figure 3:** A composite solar image on 14 February 2004, corrected for limb-darkening, with labels and arrows showing the corresponding solar regions, similar to a projection drawing.

**Table 1**

| DD | MON | YYYY | IN | IR | IG | IA | IQ | DT | DN | DR | DG | DQ |
|----|-----|------|----|----|----|----|----|----|----|----|----|-----|
| 7  | Jan | 2003 | 29 | 99 | 85 | 925 | G | -90 | 59 | 109 | 60 | G |
| 13 | Jan | 2003 | 35 | 135 | 121 | 1149 | E | -2 | 56 | 136 | 97 | E |
| 16 | Jan | 2003 | 14 | 114 | 121 | 661 | G | 39 | 13 | 93 | 97 | P |
| 22 | Jan | 2003 | 15 | 85 | 85 | 463 | G | 9 | 20 | 80 | 72 | G |
| 24 | Jan | 2003 | 19 | 79 | 72 | 663 | F | 8 | 24 | 84 | 72 | F |
| 27 | Jan | 2003 | 19 | 99 | 97 | 417 | G | 3 | 19 | 79 | 72 | F |
| 28 | Jan | 2003 | 19 | 99 | 97 | 483 | F | 4 | 22 | 102 | 97 | F |
| 29 | Jan | 2003 | 21 | 81 | 72 | 509 | G | -58 | 23 | 83 | 72 | P |
| 30 | Jan | 2003 | 10 | 70 | 72 | 265 | F | -3 | 14 | 64 | 60 | F |
| 31 | Jan | 1003 | 5 | 45 | 48 | 170 | F | 20 | 7 | 47 | 48 | F |
| 3  | Feb | 2003 | 10 | 30 | 24 | 475 | F | -1 | 18 | 38 | 24 | F |
| 6  | Feb | 2003 | 23 | 83 | 72 | 375 | G | 13 | 19 | 69 | 60 | F |
| 7  | Feb | 2003 | 23 | 93 | 85 | 333 | G | 3 | 29 | 99 | 85 | G |
| 10 | Feb | 2003 | 15 | 85 | 85 | 284 | G | -8 | 19 | 89 | 85 | F |
| 19 | Feb | 2003 | 8 | 38 | 36 | 407 | F/G | 4 | 10 | 40 | 36 | F |
| 27 | Feb | 2003 | 5 | 35 | 36 | 196 | E | -25 | 6 | 36 | 36 | G |
| 28 | Feb | 2003 | 6 | 36 | 36 | 610 | E | -36 | 8 | 38 | 36 | E |
| 3  | Mar | 2003 | 17 | 87 | 85 | 913 | G | -13 | 16 | 56 | 48 | F |
| 6  | Mar | 2003 | 14 | 64 | 60 | 695 | F/G | 4 | 13 | 53 | 48 | G |

**Table 1:** Direct comparison of spot counts R-number from ISOON; (first letter designated as I; for image) and drawing (first letter D, designated as drawing). The date designations are apparent. For ISOON digital images: IN, IR, IG, IA and IQ represent number of individual spots, sunspot number, group number, area (micro hemispheres) and image quality (Good, Excellent, Fair or Poor), respectively. DT is the time difference between a digital image and a drawing, in minutes. For the drawing images: DN. DR, DG, and DQ represent number of individual spots, sunspot number, group number and seeing quality (Excellent, Good, Fair, or Poor), respectively. Since the same observer (author Tim Henry) has to record both data, the time difference between the two processes is shown, in the column DT. Group number IG is determined by 12.08 × number of groups (see Hoyt and Schatten, 1998).

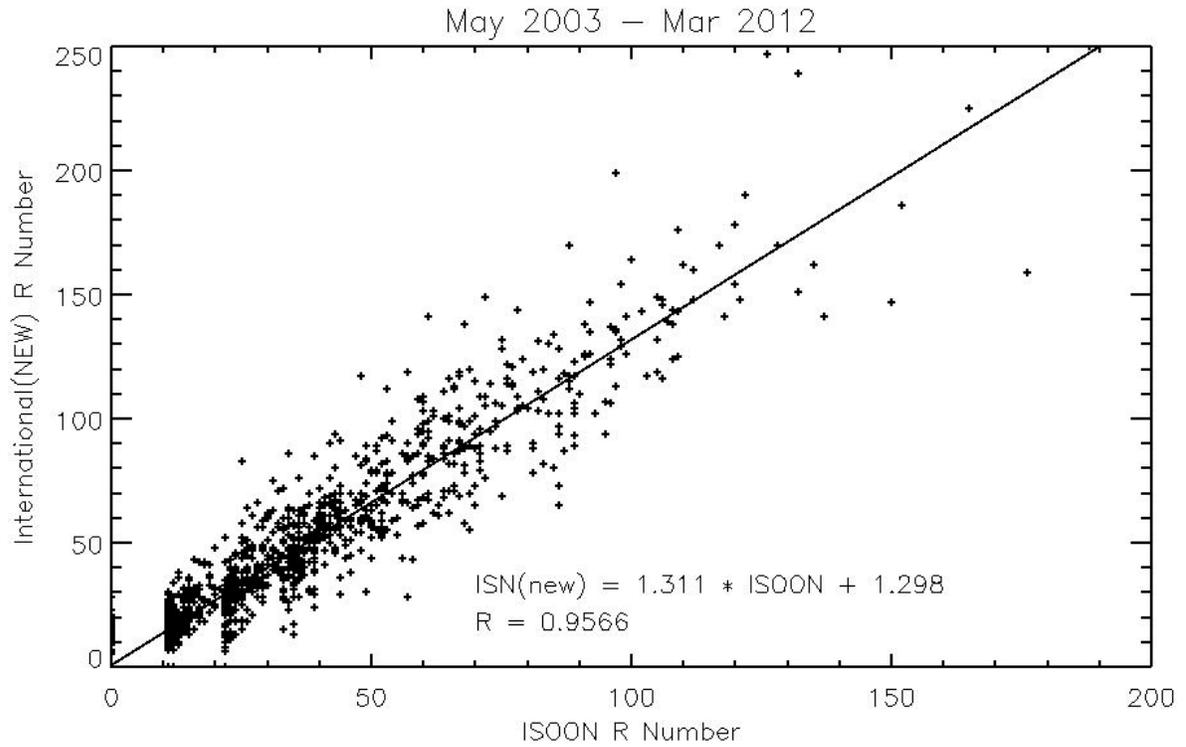

**Figure 4:** The International Sunspot Number *versus* the ISOON *R*-number during the study period described in the text. The data are plotted only on days when ISOON digital imagery was available.

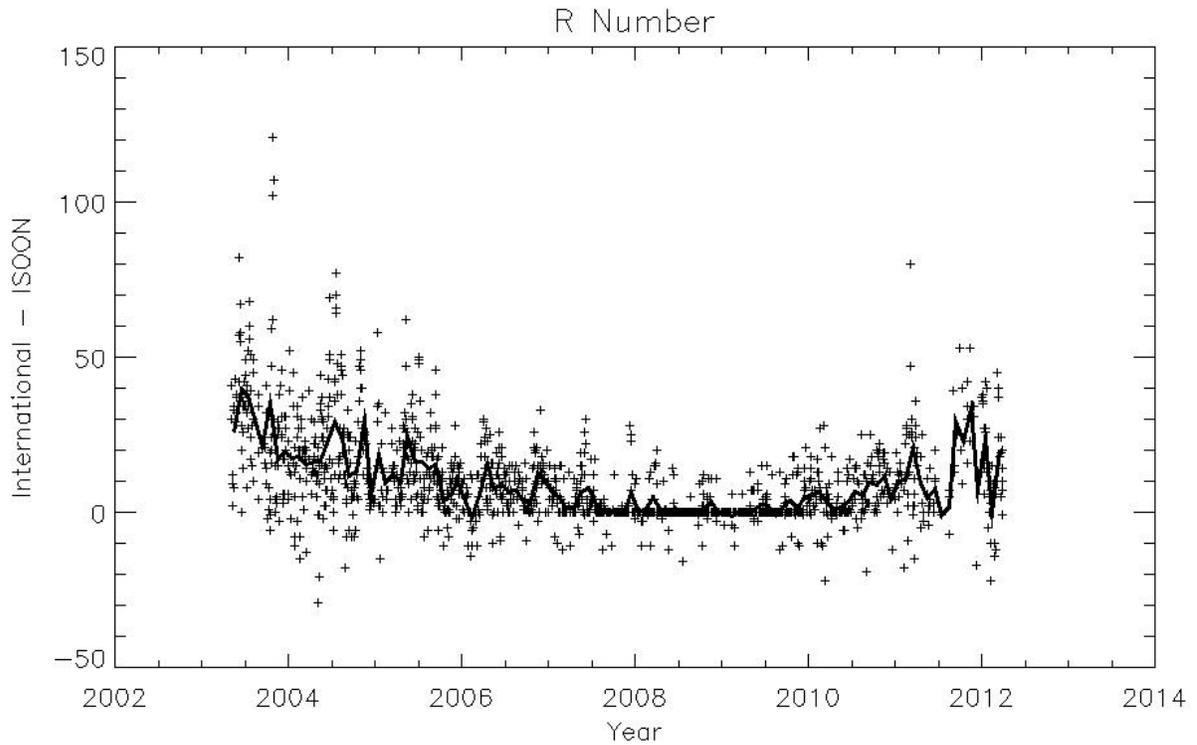

**Figure 5:** A nearly decade-long comparison of new ISN to ISOON measures, as a difference of the two measures. The continuous line shows a monthly average and the dotted points are the individual numbers.   The ISN overestimates the number of sunspots during high solar activity years.

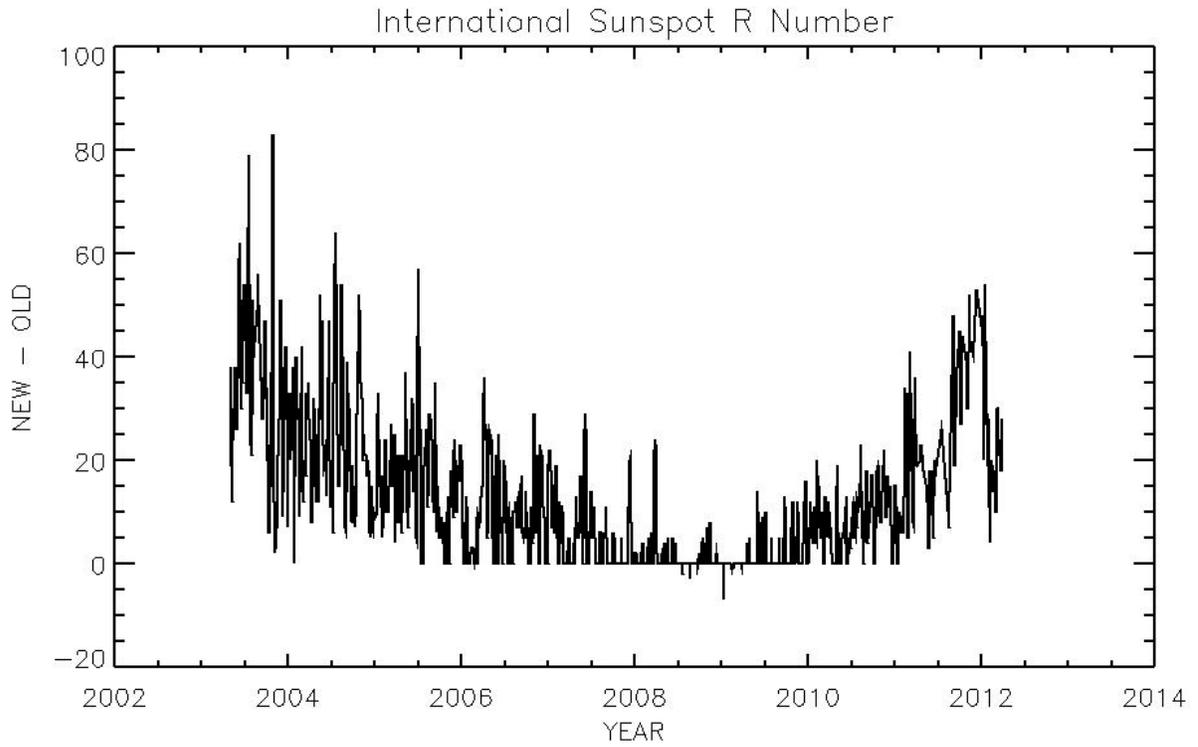

**Figure 6:** Difference between NEW ISN *versus* the OLD ISN (see text) during the period of ISOON observations.

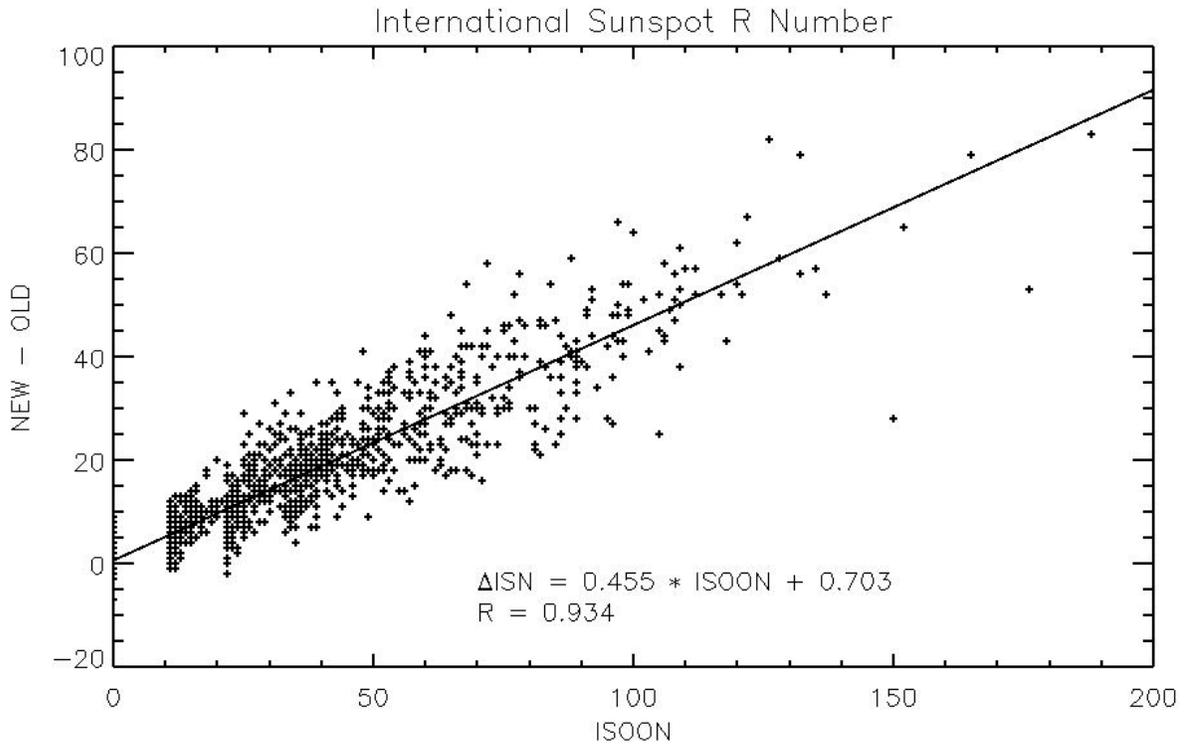

**Figure 7:** Scatter plot of the difference between NEW ISN and OLD ISN *versus* ISOON number counts, during the period of ISOON observations (2002 – 2012) (see text).

Table 2

| Year | M (slope) | B (Intercept) | R (Correlation) | NP (# points) |
|---|---|---|---|---|
| 2003 | 1.259 | 10.036 | 0.920 | 118 |
| 2004 | 1.288 | 5.580 | 0.878 | 209 |
| 2005 | 1.276 | 3.984 | 0.912 | 98 |
| 2006 | 1.199 | 3.330 | 0.917 | 188 |
| 2007 | 1.254 | 0.853 | 0.918 | 98 |
| 2008 | 1.121 | 0.524 | 0.898 | 211 |
| 2009 | 1.115 | 0.844 | 0.866 | 195 |
| 2010 | 1.082 | 3.990 | 0.879 | 161 |
| 2011 | 1.181 | 4.928 | 0.921 | 85 |
| 2012 | 1.198 | 2.246 | 0.847 | 44 |
| ALL | **1.311** | **1.298** | **0.957** | **1606** |

**Table 2:** Yearly, straight-line least-square fits of the International R-number to the ISOON *R*-number.  *M* is the slope, *B* the intercept, *R* the correlation coefficient, and *NP* is the number of points in each year.   Notice that the ten-year mean correlation coefficient is high because there are a disproportionate number of points (422 days) during which there were no sunspots during the solar minimum.

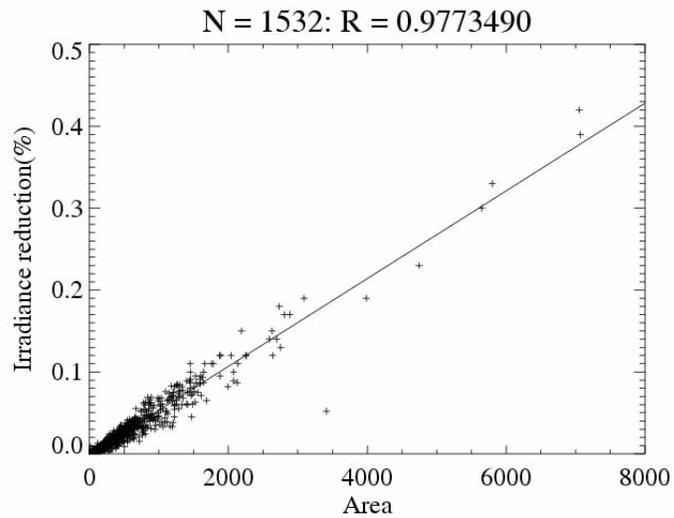

**Figure 8:** Irradiance reduction in sunspots compared to the area. Notice the high correlation and proportionality.   Extremities in the data such as the 2003 Halloween sunspots (28 – 29 October 2003) are at the top right. Note: The area measure accounts for foreshortening, while the irradiance measure has no such correction, by definition.

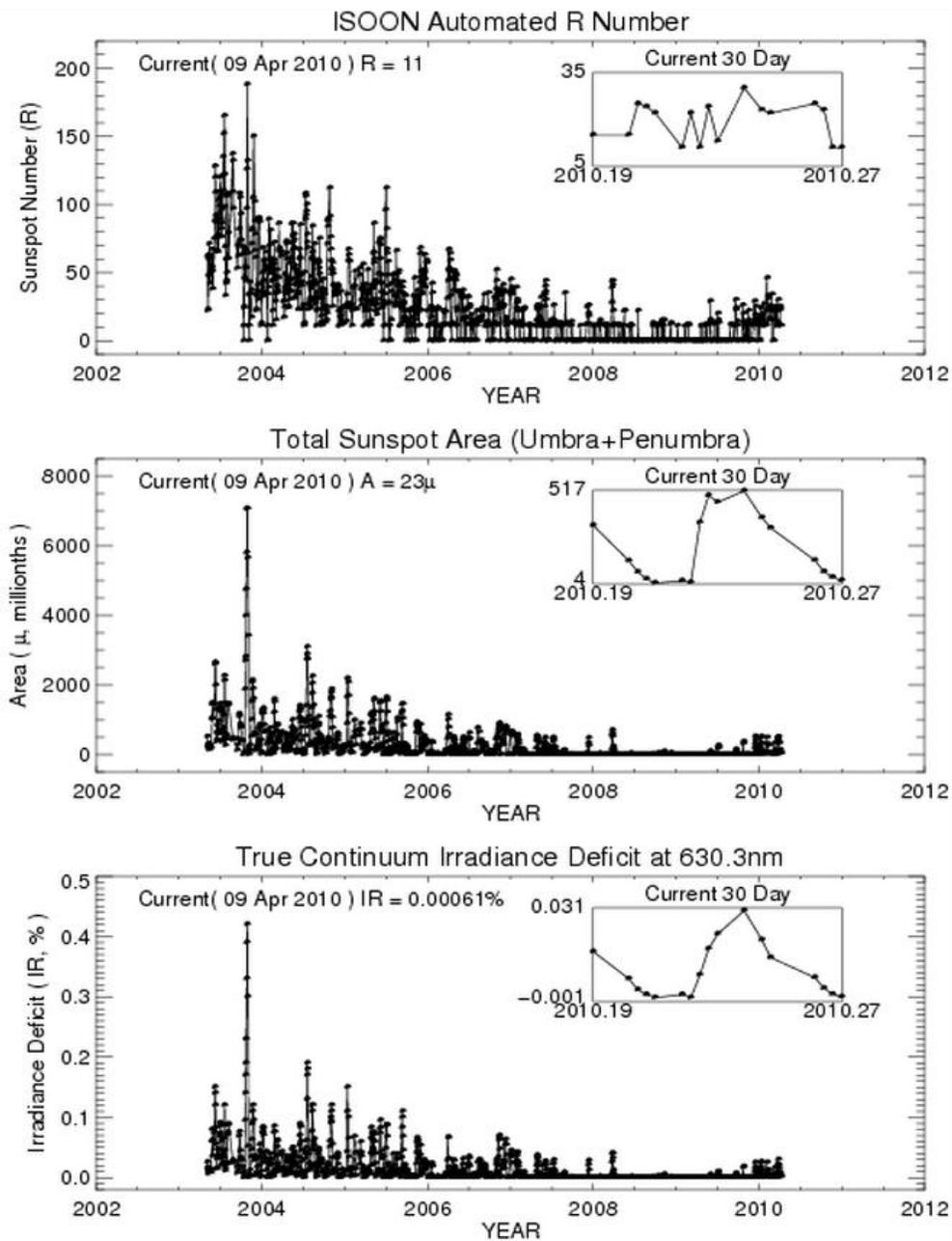

**Figure 9:** Automated Sunspot Number [R], sunspot areas, and irradiance reductions from 2003 – 2009. The inset in each plot demonstrates the prior 30-day variations of each of the quantities, reckoned from the current day (which in this case was 09 April 2010).

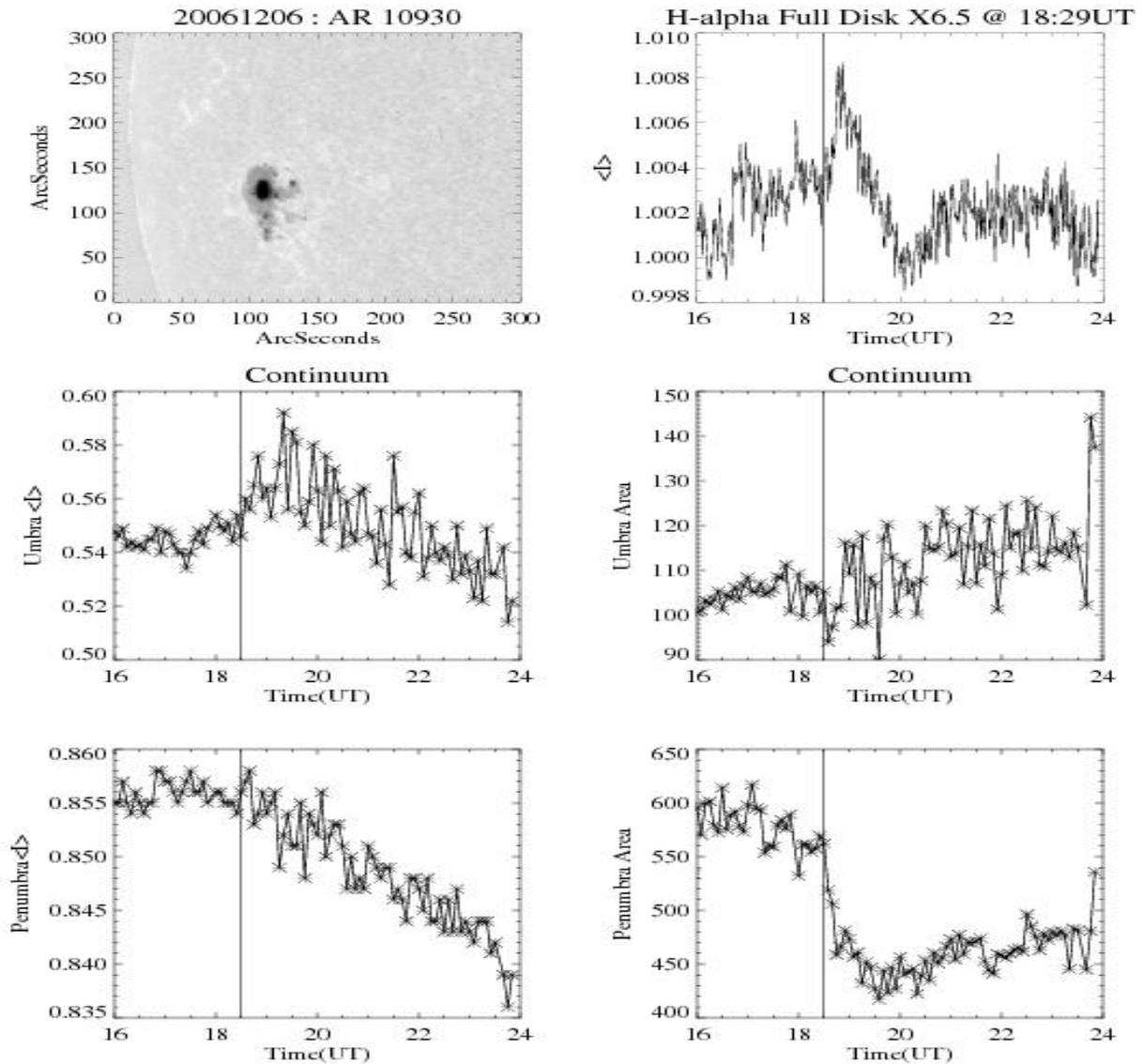

**Figure 10:** Changes in sunspot penumbral and umbral areas and intensities during the course of a day when a large flare occurred. The panels show the evolution of corresponding areas and intensities for the X6.5 flare of AR 10930 on 06 December 2006. The corresponding mean Hα intensity for the flare, averaged over the active region is shown on the top-right panel to help identify the start of the flare times. Such changes cannot be discerned simply by using a sunspot number.